# On the Complexity of Deciding Degeneracy in Games

Ye Du

Department of Electrical Engineering and Computer Science
University of Michigan, Ann Arbor, USA
duye@umich.edu

October 29, 2018

**Abstract**

We show that it is NP-Complete to decide whether a bimatrix game is degenerate and it is Co-NP-Complete to decide whether a bimatrix game is nondegenerate.

## 1 Introduction

Game theory is a subject to study and predict behaviors of rational decision makers. Nash [1] in 1950 showed that under mild conditions, a noncooperative game always has a Nash equilibrium. However, Nash's proof of the existence of Nash equilibrium is nonconstructive. People have devoted to studying the computation of Nash equilibria since then. On the complexity side, it has been shown that it is PPAD-complete to compute a Nash equilibrium, even for bimatrix games [6, 5]. On the algorithmic side, a nice property about bimatrix games is that as long as the payoff matrices are rational, the equilibria must be rational, too. Lemke and Howson [3] designed a combinatorial algorithm that can not only compute equilibria of a bimatrix game but also prove the existence of equilibria constructively. Nevertheless, just like the Simplex algorithm for linear programming, Lemke and Howson's algorithm can fail on degenerate games. As linear programming, degeneracy in games can be solved by perturbation techniques [2].

In this paper, we will investigate the computational complexity of deciding degeneracy in bimatrix games. We will show that it is NP-Complete to decide whether a bimatrix game is degenerate while it is Co-NP-Complete to decide whether it is nondegenerate.

## 2 Preliminaries

### 2.1 Definitions

Bimatrix games are the simplest cases of normal form games. In a bimatrix game, there are two players called the *row player* and the *column player* with pure strategy spaces $\mathcal{R}$ and $\mathcal{C}$ respectively. We can use two matrices $A$ and $B$ to represent the payoff matrices to the row player and the column player when they play different combinations of pure strategies. Specifically, $A_{ij}$ is the payoff to the row player when he plays its $i$th pure strategy while the column players plays its $j$th pure strategy. A mixed strategy is a probability distribution on the strategy space. Let $\Delta_m$ and $\Delta_n$ be the mixed strategy spaces of the row and the column player.

**Definition 1.** (Nash equilibrium) Given the payoff matrices $(A, B)$ of the row and column players, a strategy profile $(x^*, y^*)$ is a Nash equilibrium iff $\forall x \in \Delta_m$, $(x^*)^T A y^* \geq x^T A y^*$ and $\forall y \in \Delta_n$, $(x^*)^T B y^* \geq (x^*)^T B y$.

A closely related concept to Nash equilibrium is *best response condition*.

**Definition 2.** [2] (Best Response condition) Let $x$ and $y$ be the strategies for the row and column player respectively. $x$ is the best response to $y$ iff

$$x_i > 0 \Rightarrow (Ay)_i = u = \max\{(Ay)_k | k \in M\}$$

Now are ready to define degeneracy in games.

**Definition 3.** [2] (Nondegenerate) A bimatrix game is *nondenegerate* if there is no mixed strategy of size $k$ has more than $k$ pure best responses.

Otherwise, if the above condition is violated, the game is called *degenerate*.

## 2.2 Degeneracy in Linear Programming and in Bimatrix Games

In a linear system,

$$\begin{cases} Ax = b \\ x \geq 0 \end{cases}$$

where $A$ is a matrix of dimension $m \times n$ and rank $m$. This system is said to be *degenerate*, if there exists a basis $B$ such that at least one component in the vector $B^{-1}b$ is zero. In the next, we will study the relationship between degeneracy in linear programming and bimatrix games. If the linear system is degenerate, $b$ can be expressed as a linear combination of at most $m - 1$ columns of A. It is [4] shown that deciding whether a linear programming is degenerate is NP-Complete.

We want to discuss the relationship between *degeneracy in linear programming* and *degeneracy in games*. Let $A = \begin{pmatrix} 1 & 2 \\ 0 & 1 \end{pmatrix}$ and $b = (2, 1)^T$. It is easy to see that this linear system is degenerate. However, if $A$ is the payoff matrix of the row (the column) players, the game is nondegenerate. Reversely, $A = \begin{pmatrix} 1 & 1 \\ 0 & 1 \end{pmatrix}$ and $b = (2, 1)^T$. If $A$ is the payoff matrix in a bimatrix game, the game is degenerate. However, the linear system is nondegenerate. Therefore, degeneracy in linear programming does not imply degeneracy in games; *vice versa*.

## 3 The Main Theorem

**Theorem 1.** *It is NP-complete to decide whether a bimatrix game is degenerate.*

**The construction:** We will reduce the 3-SAT to our problem. Let $f$ be a 3-SAT formula $f = c_1 \wedge c_2 \wedge ... \wedge c_n$, where each clause $c_i = l_{i1} \vee l_{i2} \vee l_{i3}$ contains three literals. Each literal $l_{iw}$ is either a positive variable $x_h$ or its negation $\overline{x_h}$. Let $c_i^k$ denote the assignment of the true values of the three literals in $c_i$ as the binary representation of the integer $k$, where $k \in [1..7]$. For example, the binary representation of 5 is 101. Thus, $c_i^5$ represents the assignment such that $l_{i1} = 1, l_{i2} = 0$ and $l_{i3} = 1$.

The strategy space of the column player is $\mathcal{C} = \{c_i^k | i \in [1..n] \text{ and } k \in [1..7]\}$. And the strategy space $\mathcal{R}$ of the row player is $\mathcal{C} \bigcup \{f\} \bigcup \{c_i^p c_j^q | c_i^p \text{ and } c_j^q \text{ are conflicting}\}$, where $f$ is a special strategy. Two clauses assignments $c_i^p$ and $c_j^q$ are conflicting iff there are two literals $l_{iw}$ in $c_i$ and $l_{js}$ in $c_j$ that are corresponding to the same variable $x_h$ and $c_i^p$ and $c_j^q$ make conflicting assignments to $x_h$. For example, let $c_1 = x_1 \vee \overline{x_2} \vee x_3$ and $c_2 = x_2 \vee x_4 \vee x_5$. $c_1^5$ will make $\overline{x_2} = 0$, which assigns $x_2 = 1$ and $c_2^3$ will assign $x_2 = 0$. Thus, $c_1^5$ and $c_2^3$ are conflicting clauses assignments. Note that the order of $c_i^p$ and $c_j^q$ does not matter, i.e., the pure strategy $c_i^p c_j^q$ implicitly equals the pure strategy $c_j^q c_i^p$.

Given the strategy space of the row and column players, the payoff function $r$ for the row player is

1. $r(c_i^p, c_h^k) = 1$ if $i = h$ and $p = k$; otherwise $r(c_i^p, c_h^k) = 0$.

2. $\forall h \in [1..n]$ and $k \in [1..7]$, $r(f, c_h^k) = 1/n$.

3. Suppose we have $D$ conflicting clauses assignments $c_i^p c_j^q$ and let $g$ be an ordering of the elements in $\{c_i^p c_j^q | c_i^p \text{ and } c_j^q \text{ are conflicting}\}$. Thus, for the $d$th conflicting clauses assignments $c_i^p c_j^q$, $r(c_i^p c_j^q, c_h^k) = \frac{1}{2} + 3^d \epsilon$ if $i = h$ and $p = k$ or $j = h$ and $q = k$; otherwise $r(c_i^p c_j^q, c_h^k) = 0$. Here $\epsilon$ is a small positive number whose values will be fixed later. Note that $D \in \theta(n^2)$.

The following matrix illustrates the payoff function of the row player.

|  | $c_1^1$ | $c_1^2$ | ... | $c_1^7$ | $c_2^1$ | $c_2^2$ | ... | $c_2^7$ | ......... | $c_n^1$ | $c_n^2$ | ... | $c_n^7$ |
|---|---|---|---|---|---|---|---|---|---|---|---|---|---|
| $c_1^1$ | 1 | | | | | | | | | | | | |
| $c_1^2$ | | 1 | | | | | | | | | | | |
| . | | | . | | | | | | | | | | |
| . | | | | | | | | | | | | | |
| $c_1^7$ | | | | 1 | | | | | | | | | |
| $c_2^1$ | | | | | 1 | | | | | | | | |
| . | | | | | | . | | | | | | | |
| . | | | | | | | | | | | | | |
| $c_2^7$ | | | | | | | | 1 | | | | | |
| . | | | | | | | | | . | | | | |
| . | | | | | | | | | | | | | |
| $c_n^1$ | | | | | | | | | | 1 | | | |
| . | | | | | | | | | | | . | | |
| . | | | | | | | | | | | | | |
| $c_n^7$ | | | | | | | | | | | | | 1 |
| $f$ | $\frac{1}{n}$ | $\frac{1}{n}$ | $\frac{1}{n}$ | $\frac{1}{n}$ | $\frac{1}{n}$ | $\frac{1}{n}$ | $\frac{1}{n}$ | $\frac{1}{n}$ | $\frac{1}{n}$ | $\frac{1}{n}$ | $\frac{1}{n}$ | $\frac{1}{n}$ | $\frac{1}{n}$ |
| $c_1^1 c_1^2$ | $\frac{1}{2} + \epsilon$ | $\frac{1}{2} + \epsilon$ | | | | | | | | | | | |
| . | | | | | | | | | | | | | |
| $c_i^p c_j^q$ | | | | | | $\frac{1}{2} + 3^d \epsilon$ | | | | | $\frac{1}{2} + 3^d \epsilon$ | | |
| . | | | | | | | | | | | | | |
| $c_n^6 c_n^7$ | | | | | | | | | | | | $\frac{1}{2} + 3^{D-1} \epsilon$ | $\frac{1}{2} + 3^{D-1} \epsilon$ |

*Proof.* Given the mixed strategy $y$ of the column player with support $k$, we can multiply $y$ with the payoff matrix of the row player. Then, if the number of best responses of the row player is more than $k$, the game is degenerate. Thus, deciding degeneracy is in NP. We will reduce the 3-SAT problem to the problem of deciding degeneracy in games as shown in above. It is obvious that the size of the payoff matrix is in a polynomial of $n$ and each entry of the matrix can be represented by a polynomial number of bits. Therefore, the construction can be done in polynomial time.

If there is a satisfying assignment to the variables of the formula $f$, the assignment of each clause $c_i^k$ must be in the form of $c_i^k$, where $k \in [1..7]$. We call such a corresponding strategy $c_i^k$ *active pure strategy*. We will set the mixed strategy $y$ of the column player to be $\frac{1}{n}$ on those active pure strategies and 0 at any other pure strategies. Thus, the support or $y$ is $n$. Given such a $y$, it is easy to see, for the row player, all the active pure strategies and $f$ are the best responses to $y$, i.e., the number of best responses is $n+1$. Therefore, if $f$ is satisfiable, the game is degenerate.

In the next, we prove the reverse direction. Suppose the game is degenerate. Again let $y$ the strategy of the column player, $u$ be the maximum payoff to the row player given $y$ and $\mathcal{S}$ be the support set of $y$. Assume that the support set $\mathcal{S}$ can be represented as the union of two subsets $\mathcal{M} = \{c_i^k | y[c_i^k] = u\}$ and $\mathcal{N} = \{c_i^k | y[c_i^k] < u\}$. We will show that the set $\mathcal{N}$ is empty in the next.

Given $y$, we will count the number of best responses of the row players respect to $y$. First of all, for strategy $c_i^k \in \mathcal{M}$, its corresponding pure strategy for the row player is one of the best responses. Secondly, the payoff to the strategy $f$ is $\frac{1}{n}$. For any pure strategy $c_i^p c_j^q$, if both $c_i^p \in \mathcal{M}$ and $c_j^q \in \mathcal{M}$, its payoff will exceed $u$, which contradicts the maximum payoff assumption of $u$. Thus, at most one of $c_i^p$ and $c_j^q$ can belong to $\mathcal{N}$. Now assume that $c_{i_1}^{p_1} \in \mathcal{N}$ and the corresponding pure strategy $c_{i_1}^{p_1} c_{j_1}^{q_1}$ of the row player is one of the best responses, while $c_{j_2}^{q_2} \in \mathcal{N}$ and the corresponding pure strategy $c_{i_2}^{p_2} c_{j_2}^{q_2}$ of the row player is also one of the best responses. Moreover, assume that $c_{i_1}^{p_1} c_{j_1}^{q_1}$ is at the $d_1$th row of the payoff matrix, $c_{i_2}^{p_2} c_{j_2}^{q_2}$ is at the $d_2$th row and $c_{i_1}^{p_1} c_{j_2}^{q_2}$ is the $d$th row, where $d_1 < d_2 < d$. Thus, we claim that the payoff to $c_{i_1}^{p_1} c_{j_2}^{q_2}$ will be more than $u$ when $\epsilon$ is small enough. Note that, by the assumption of best response condition,

$$(\frac{1}{2} + 3^{d_1}\epsilon)(y[c_{i_1}^{p_1}] + y[c_{j_1}^{q_1}]) = u$$

, we know that $y[c_{j_1}^{q_1}] \leq u$, thus $y[c_{i_1}^{p_1}] \geq \frac{1-2*3^{d_1}\epsilon}{1+2*3^{d_1}\epsilon}u$. Similarly, we can get $y[c_{j_2}^{q_2}] \geq \frac{1-2*3^{d_2}\epsilon}{1+2*3^{d_2}\epsilon}u$. Set $\epsilon = \frac{1}{6*3^{2D}}$, the payoff to $c_{i_1}^{p_1} c_{j_2}^{q_2}$ would be

$$\begin{aligned}
(\frac{1}{2} + 3^d\epsilon)(\frac{1-2*3^{d_1}\epsilon}{1+2*3^{d_1}\epsilon} + \frac{1-2*3^{d_2}\epsilon}{1+2*3^{d_2}\epsilon})u &> (1+2*3^d\epsilon)\frac{1-2*3^{d_2}\epsilon}{1+2*3^{d_2}\epsilon}u \\
&\geq (1+6*3^{d_2}\epsilon)\frac{1-2*3^{d_2}\epsilon}{1+2*3^{d_2}\epsilon}u \\
&= \frac{1+4*3^{d_2-2D} - 12*3^{2(d_2-2D)}}{1+2*3^{d_2-2D}}u \\
&> u
\end{aligned}$$

Thus, for any pure strategy $c_i^p c_j^q$, if it is one of the best responses respect to $y$, either $c_i^p$ or $c_j^q$ will be in $\mathcal{N}$ and it can not appear in any best response pure strategy $c_{i'}^{p'} c_{j'}^{q'}$, which has smaller order than $c_i^p c_j^q$. In other words, each best response pure strategy $c_i^p c_j^q$ must consume one element in $\mathcal{N}$, which is not shared with other best response conflicting clauses strategies.

Now, we can count the number of best responses respect to $y$. First of all, we know that $u \geq \frac{1}{n}$, since the payoff to the strategy $f$ is always $\frac{1}{n}$. If $u > \frac{1}{n}$, the number of best responses is $|\mathcal{M}| + |\mathcal{N}| = |\mathcal{S}|$. Thus, the game is nondegenerate. If $u = \frac{1}{n}$, let $n'$ be the number of best responses conflicting clauses strategies. We will show that $n' < |\mathcal{N}|$. Otherwise, assume $n' = |\mathcal{N}|$. ($n'$ can not be greater than $|\mathcal{N}|$ since each best response pure strategy $c_i^p c_j^q$ must consume one nonshared element in $\mathcal{N}$). Since for each best response conflicting clause strategy $c_i^p c_j^q$, either $c_i^p$ or $c_j^q$ will be in $\mathcal{N}$. W.L.O.G., we assume $c_i^p \in \mathcal{N}$. We also know that $\frac{1-2*3^D\epsilon}{1+2*3^D\epsilon}\frac{1}{n} \leq y[c_i^p] < \frac{1}{n}$. Thus, there must be a strategy $c_i^k \in \mathcal{N}$ such that

$$\begin{aligned}
0 < y[c_i^k] &\leq 1 - \frac{|\mathcal{M}|}{n} - \frac{1 - 2*3^D \epsilon}{1 + 2*3^D \epsilon} \frac{|\mathcal{N}| - 1}{n} \\
&\leq \frac{|\mathcal{M}| + |\mathcal{N}| - 1}{n} - \frac{|\mathcal{M}|}{n} - \frac{1 - 2*3^D \epsilon}{1 + 2*3^D \epsilon} \frac{|\mathcal{N}| - 1}{n} \\
&\leq \frac{|\mathcal{N}| - 1}{n} \frac{4*3^D \epsilon}{1 + 2*3^D \epsilon}
\end{aligned}$$

we know that when $\epsilon = \frac{1}{6*3^{2D}}$,

$$\frac{|\mathcal{N}| - 1}{n} \frac{4*3^D \epsilon}{1 + 2*3^D \epsilon} < \frac{1 - 2*3^D \epsilon}{1 + 2*3^D \epsilon} \frac{1}{n}$$

It implies that the strategy $c_i^k$ can not appear in any best response conflicting clause strategy. Thus $n' < |\mathcal{N}|$. Moreover, $f$ is one of the best response when $u = \frac{1}{n}$. In all, when $u = \frac{1}{n}$, the number of the best responses for the row player respect to $y$ can not be more than $|\mathcal{N}| + |\mathcal{M}|$. Thus the game in nondegenerate. Since we assume that the game is degenerate, contradiction occurs. Therefore, $\mathcal{N}$ is empty. Equivalently, we know that $\mathcal{S} = \mathcal{M}$. We further claim that $u = \frac{1}{n}$. Otherwise, if $u > \frac{1}{n}$, $f$ can not be one of the best responses and any conflicting clauses strategy can not be either. The game is still degenerate. Thus, $u = \frac{1}{n}$. Moreover, we know that for any $c_i^p$ and $c_i^q$, both $y[c_i^p]$ and $y[c_i^q]$ can equal to $\frac{1}{n}$ simultaneously. Otherwise, the payoff to conflicting clauses strategy $c_i^p c_i^q$ is more than $\frac{1}{n}$. Reversely, for every clauses $c_i$, there must be a $k$ such that $c_i^k \in \mathcal{S}$ by the pigeon hole principle. Thus, each clause is satisfied. Similarly, no two conflicting clauses $c_i^p$ and $c_j^q$ can be in the support set $\mathcal{S}$ of $y$. Therefore, the clauses assignment $(c_1^{k_1}, c_2^{k_2}, ..., c_n^{k_n})$ corresponding to $\mathcal{S}$ gives a satisfying assignment to the formula $f$.

Finally, we have shown that the formula $f$ is satisfiable iff the game is degenerate. Thus, deciding degeneracy is NP-Complete. □

A straightforward corollary of Theorem 1 is that,

**Corollary 1.** It is Co-NP-Complete to decide whether a game is nondegenerate.

In the next, we want to study a special class of bimatrix games: *win-lose games*. In *win-lose* bimatrix games, the payoff value is either 0 or 1. The following is the necessary and sufficient condition for a win-lose bimatrix game to be nondegenerate.

**Theorem 2.** *For a win-lose bimatrix game, it is nondegenerate iff for the row player, every column of its payoff matrix A has at most 1 nonzero element while for the column player, every row of its payoff matrix B has at most 1 nonzero element, too.*

*Proof.* We prove our claim only for the row player and the proof for the column player follows symmetrically. Suppose a win-lose game is nondegenerate, each column of matrix $A$ can not have more than 1 nonzero item. Otherwise, if the column player plays a pure strategy of that column, the row player has more than 1 best responses. Thus, the game is degenerate. Contradiction occurs. On the other hand, if each column of matrix $A$ has at most 1 nonzero element, for each strategy $y$ of the column player, the number of nonzero elements in the vector $Ay$ can not be more than the size of support in $y$. Thus, the game is nondegenerate. □

## 4  Conclusion and Future Works

In this paper, we show that it is NP-Complete to decide whether a bimatrix game is degenerate while it is Co-NP-Complete to decide whether it is nondegenerate. Although degeneracy in bimatrix games as well as in linear programming can be solved by perturbation techniques. It is unclear that how the perturbations affect the computational efficiency of the Lemke-Howson algorithm. It is worthwhile to further investigate the impact of degeneracy in the computation of Nash equilibria. Moreover, the design of PTAS for Nash equilbria is widely open. In all, we believe that the study of computations of Nash equilibria is a challenging and interesting area.